\documentclass[preprint, amsmath, amssymb]{revtex4}
\usepackage{graphicx}
\usepackage{dcolumn}
\usepackage{bm}
\newcommand{\beq}{\begin{equation}}
\newcommand{\enq}{\end{equation}}

\newcommand {\lav}{\langle\!\langle}
\newcommand {\rav}{\rangle\!\rangle}
\newcommand{\bra}{\langle}
\newcommand{\ket}{\rangle}

\begin{document}

\title[Author guidelines for IOPP journals]{
Anharmonicity and asymmetry of Landau levels for a two-dimensional electron gas}

\author{Stephane Bonifacie,  Christophe Chaubet\footnote[3]{To whom correspondence should be addressed (chaubet@ges.univ-montp2.fr)}, Benoit Jouault and Andre Raymond }
\affiliation{Groupe d'Etude des Semiconducteurs, UMR CNRS 5650, Universit\'e Montpellier II, 34095 Montpellier cedex France.} 

\begin{abstract}
We calculate the density of states of a two dimensional electron gas located at the interface of a GaAlAs/GaAs heterojunction. The disorder potential which is generally created by a single doping layer behind a spacer, is here enhanced by the presence of a second delta doped layer of scatterers which can be repulsive or attractive impurities. We have calculated the density of states by means of the Klauder's approximation, in the presence of a magnetic field of arbitrary strength. At low field, either band tails or impurity bands are created by attractive potentials, depending on the impurity concentration. At higher field, impurity bands are created by both repulsive and attractive potentials. The effects of such an asymmetrical density of states on the transport properties in the quantum Hall effect regime, are also discussed.

\end{abstract}


\maketitle

\section{Introduction}
Although the calculation of the density of states (DOS) for a two dimensional electron gaz (2DEG) is an old problem, it has gained interest in the last decade because of new types of heterostructures grown by several groups~\cite{haug,merkt,bonifacie,meulen,zhao,halsall,Wongmanerod,holtz,hayne,kukushkin}. The heterostructures considered in this article are GaAlAs/GaAs heterojunctions. They have, as usual, a delta doped layer behind the spacer to provide electrons in the 2D channel. Besides, they have another delta doped layer (Si donors or Be acceptors) inside the confining well of conducting electrons. Transport properties in high magnetic field have been studied on this kind of samples by R.J. Haug et al.~\cite{haug}. These authors observed a shift of the quantum Hall plateaus (for a review on the quantum Hall effect, see ref.~\cite{prange}). Merkt et al.~\cite{merkt} and Bonifacie et al.~\cite{bonifacie} have performed cyclotron resonance experiments on these structures. Two main phenomena have been clearly observed. The first one concerns a new cylotron line, called disorder mode~\cite{merkt}, which is observed only for the acceptors-doped structures, and might be attributed to the coupling between the Landau harmonic oscillators and the background oscillator created by the presence of Be-atoms in the cristal. The second concerns a set of additional lines attributed to the energy levels created by the acceptors background~\cite{bonifacie}. Kubisa and Zawadzki~\cite{kubisa} pointed out that repulsive potentials create localized states when a magnetic field is present, by the combined effect of the Lorentz force and the electrical confinement. Although localized states created by repulsive potentials are at higher energy than the corresponding free Landau states, they can still trap electrons when the cyclotron orbit is shrinked by increasing magnetic field. These authors calculated the binding energies of magneto-acceptors, and showed that, contrarily to donors, acceptors states disappear at zero magnetic field.\\

In this paper we calculate the DOS of heterojunctions having one or two delta doped layers. We use the Green function formalism and perform the calculations using a numerical method proposed earlier by Klauder\cite{klauder}: the so-called fifth Klauder's approximation (KVA)~\cite{klauder}. The KVA was already employed by Ando~\cite{ando1} in the case of high fields, to obtain the DOS of a two dimensional electron gas(2DEG) in presence of a single type of impurities. More recently it was employed as well by Serre~{\it et al.}~\cite{serre} and Gold ~{\it et al.}\cite{gold} at zero magnetic field, in three and two dimensions systems. The most severe limitation of the KVA is that the multiple occupancy corrections are not taken into account. Consequently the width of the impurity band is overestimated, resulting in a Mott density which is too small~\cite{gold,monecke}. Another factor of imprecision is related to our choice to not incorporate in our model, the self energy part due to the electron-electron interaction. For sake of simplicity, we focused on the general asymmetrical shape of the DOS created by a given concentration of impurities. Several approximations for this interaction have been discussed in the litterature~\cite{afs}, but following  ref.~\cite{gold}, the self-energy part $\hat{\Sigma}_{ee}$ only produces a rigid energy shift of the DOS. Besides, these approximations do not hold at high magnetic field, when the filling factor is close to one and electron-electron interaction strongly increases the Lande g-factor. On the other hand, we took into account the mixing between landau Levels (LLs) which has never been done before. Indeed, most microscopic theories~\cite{joynt} and calculations of the DOS~\cite{ando1,brezin,wegner,affleck} either consider only one LL, neglecting inter-LL mixing, or take into account only two or three LLs~\cite{ando2}. Consequently, the  DOS is correctly reproduced in the strong-field limit or in the low-field limit, but is not so well known for intermediate or low fields, where the strength of the random fluctuations becomes comparable to the energy separation between LLs.\\

This paper is organized as follows. In the next section, we present our theoretical treatment. We reported the results of the calculations in section III and section IV: we discuss the DOS for the reference sample having only one impurity layer in section III, while section IV is devoted to the anharmonic and asymetrical DOS. In section V, we apply this formalism to the transport properties in the QHE regime and we correctly reproduce the shift of the Hall plateaus.

\section{Formulation of the problem}
\begin{figure}
\begin{center}
\includegraphics[width=8cm]{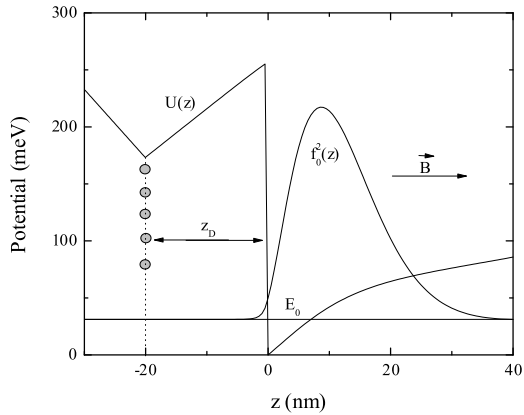}
\caption{Schematic view of the heterostructure and the  $f_0(z)$ wavefunction. A delta doped layer is represented here, located at abscissa $z_0$. A magnetic field is applied along the growth direction.}
\label{fig:figure1}
\end{center}
\end{figure}

The 2D channel is considered as a homogeneous medium with a dielectric constant $\kappa$ and the charge carriers are noninteracting electrons with an effective mass $m^*$ and a charge $e$. The electrons are free in a bidimensional quantum well created by the potential $U(z)$ of a heterojunction (see Fig.~\ref{fig:figure1}). An external magnetic field is applied in the $z$ direction. One or two delta layers of impurities are introduced at a given distance from the GaAlAS/GaAs interface. The first doping layer has a concentration comparable to the electron concentration (few $10^{11}$ cm$^{-2}$). The second layer of impurities, located at the distance $z_0$ from the interface with a concentration $N_i$, create sharp potential fluctuations, either negative (for donors) or positive (for acceptors). Its concentration is one order of magnitude lower than the first one (few $10^{10}$ cm$^-2$). 

\subsection{The single impurity problem}
The single impurity problem which was solved in Ref.~\cite{kubisa} by means of variational method, is here treated by a diagonalization procedure. This diagonalization method is used afterwards in the self consistent calculations of the Green function, for convergence purpose. 

The impurity located at point $(\mathbf{R}, z_0)$ creates a potential $v_0(\mathbf{r}, z)= e^2/( 4\pi\kappa \sqrt{ (z-z_0)^2 + (\mathbf{R}-\mathbf{r}) ^2})$. The three dimensional Hamiltonian of this system is given by
\begin{equation}
{\hat{H}}= \frac{1}{2m^*} \left( \hat{\mathbf{p}} + 
e  \hat{\mathbf{A}} \right)^2 + \hat{U} + \hat{v}_\mathrm{0}~~~, 
\end{equation}
where ${\hat{\mathbf{A}}}$ is the vector potential operator and $\hat{\mathbf{p}}=(\hat{\mathbf{p_\bot}},\hat{p_z})$ the momentum operator. To solve this equation the 3 dimensional problem is first reduced to a bidimensional one~\cite{brum}. The wavefunctions can be approximated by
\begin{equation}
\Psi(x,y,z)= \Phi(x,y) f_0 (z)~~,
\end{equation}
where $f_0(z)$ is the wavefunction of the first subband of the quantum well with an energy $E_0$:
\begin{equation}
\left( \frac{1}{2m^*} \hat{p}_z^2+ \hat{U} \right) f_0  = E_0 f_0 .
\end{equation}
The eigenenergy $E_0$ is  obtained by  a variational calculation and $f_0$ is approximated by the modified Fang-Howard trial wavefunction. Details of this method can be found in ref.~\cite{bastard}. One obtains an effective Schr\"odinger equation for the $\Phi$ wavefunction of eigenenergy $E_\bot$:
\begin{equation}
\left[
\hat{H}_\bot + 
\hat{v}_{\mathrm{bare}}
\right]  \Phi = E_\bot \Phi~~,
\end{equation}
where
\begin{equation}
\hat{H}_\bot= \frac{1}{2m^*} 
\left(\hat{\mathbf{p}}_\bot +e \hat{\mathbf{A}} \right)^2
\end{equation}
and
\begin{equation}
\hat{v}_\mathrm{bare}(q)= \frac{1}{\pi q \epsilon(q)} \int_{-\infty}^{\infty} f_0(z)^2
e^{-q|z-z_0|} dz.
\end{equation}
The screening of the bare potential by free carriers has been taken into account in ref.~\cite{price} and ref.~\cite{kubisa} by introducing the dielectric function $\epsilon(q)= 1+ \frac{ e^2 }{2 \kappa} X_0(q) h(q)$, where $X_0$ is the Linhard function in two dimensions~\cite{stern} and $h(q)$ is a form factor which is due to the electron-electron interaction in the 2DEG. The whole problem has been solved directly by numerical methods. However, one should bear in mind that in the linear response approximation, the mathematical expression of the dielectric function when a magnetic field is applied is quite different from that deduced from the familiar Lindhard function in 2D. This is because both the energy levels and the wave functions are different. Nevertheless, we focus in this article on the general asymetry of the Landau levels, and not on the screening of the bare potential by electrons. This is why we did not use a different dielectric function.

\begin{figure}
\begin{center}
\includegraphics[width=8cm]{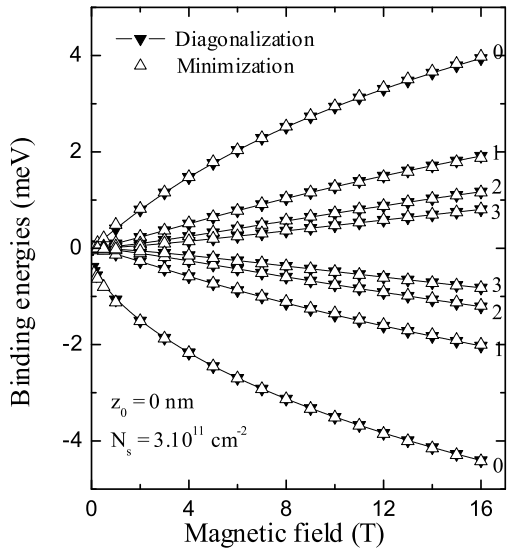}
\caption{Binding energies of magnetodonors and magnetoacceptors of the first Landau level. Parameters are given in the text. The angular momentum $m$ is equal to 0,1, 2 and 3. Black down triangles: diagonalization. White up triangles: variational calculation. At zero magnetic field, the binding energy vanishes for acceptor states but not for donors: the acceptor states exist only in presence of high enough magnetic field.}
\label{fig:figure2}
\end{center}
\end{figure}

To calculate the DOS in the KVA approximation, we have used a diagonalization procedure instead of the variational method. Now, we apply the diagonalization method to the single impurity problem, in order to demonstrate first the accuracy of our procedure. We used the basis $|N,m\ket$, $N \in \{0,1,\ldots, \infty\}$, $m \in \{-N, \ldots, \infty\}$ defined by:
\begin{equation}
\bra \mathbf{r}|N,m\ket= 
\sqrt{\frac{N!}{2\pi l^2 (N+|m|)!}} \nonumber  \times 
e^{\left( -i m \phi - \frac{r^2}{4l^2}   \right)}
\left( \frac{r^2}{4l^2} \right)^{|m|/2}
L_N^{|m|/2} \left( \frac{r^2}{2l^2}\right)
\label{phinm}
\end{equation}
where  $l= \sqrt{\hbar/eB}$ is the cyclotron length and $B$ is the magnetic field. Because of the radial symmetry of the problem, the calculations can be performed for each value of $m$ separately. In Fig.~\ref{fig:figure2} we plot the calculated binding energies of magnetodonors and magnetoacceptors states related to the first LL. The impurities are located at the interface ($z_0=0$). The parameters are the same than those used by Kubisa and Zawadzki~\cite{kubisa}: for a typical GaAs/AlGaAs heterojunction, the barrier height is $0.257$ eV, the effective mass for the AlGaAs barrier and the GaAs well are respectively $0.073m_0$ and $0.066m_0$. We took a relative dielectric constant $\kappa_r= 12.9$ throughout the entire heterostructure. The electron density and the depletion density have been taken equal to $N_s=3\times 10^{11}$ cm$^{-2}$ and $N_a=6\times 10^{11}$ cm$^{-2}$ respectively. The diagonalization has been performed in a basis containing a finite number of LLs. We have included all LL whose energy is smaller than or comparable to the binding energy $E_b$. At $B$=10T, 10 LLs have been taken into account. This allowed us to obtain a precision better than 0.1\%. At $B$=0.1T, 150 LLs have been considered to obtain a precision of $5\%$. Figure~\ref{fig:figure2} shows that there is no quantitative difference between our method and the minimization procedure used by Kubisa and Zawadzki. However the good convergence of the diagonalization procedure guarantees a reasonable calculation time for the self consistent procedure.\\

The second step of our calculation procedure was to fit the true impurity potential by a gaussian potential which allowed us to calculate faster the binding energies, and therefore to optimize our algorithm in the second part of this work: 
\begin{equation}
v(r)= \frac{V_0}{\pi d^2} exp \left(  - \frac{r^2}{d^2}\right). 
\end{equation}
where $V_0$ is the strength of the potential and $d$ is its spatial extent. This correspondance allowed us to perform the heavy self consistent calculation of the Green function procedure, developped in the following paragraph. For this reason, we have first re-calculated the binding energies of Kubisa and Zawadzki by using a diagonalisation procedure with a fitted gaussian potential. We obtained a good fit of the binding energy of Fig.~\ref{fig:figure2} by taking $V_0$= 22meV and $d$= 10.4nm. Therefore we are in the case of short-range scatterers ($d < l/ \sqrt{2N+1}$, N is the LL index)~\cite{afs} for a wide range of magnetic field (0$ <B <6$T). This corresponds to the position of impurities  $z_0=0$. When $z_0= 500 \AA$, the optimized parameters are $d=50nm$ and $V_0=0.17 meV$. When impurities are close to the electrons, the gaussian potential is short range, but its becomes long range when the impurity layer is located at several hundreds of angstroms from the interface. 

\subsection{Hamiltonian of the disordered system}
Now we consider the case of several impurities which are all located in the same delta-layer at the distance $z_0$ of the hetero-interface. The Hamiltonian of the 2DEG in the presence of these randomly distributed impurities can be written as
\begin{equation}
\hat{H} = \hat{H}_\bot + \hat{V} ,
\end{equation}
with
\begin{equation}
\hat{V}= \sum_i \hat{v}_i
\end{equation}
and
\begin{equation}
 \hat{v}_i=  \hat{v} (\mathbf{r}-\mathbf{R_i}, z_0)~.
\end{equation}
Here $\mathbf{R}_i$ are the in-plane positions of the different impurities.

\begin{figure}
\begin{center}
\includegraphics[width=8cm]{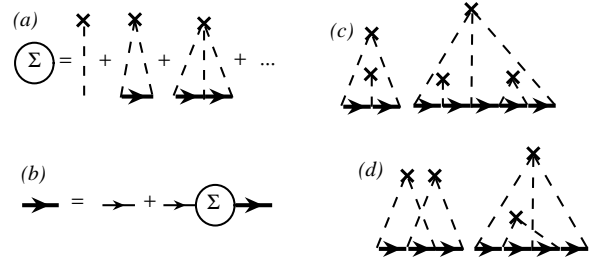}
\caption{Electron-impurity diagrams involved in the KVA. (a) summation of the diagrams of the self-energy; (b) Dyson equation; (c) examples of diagrams that are taken into consideration in the KVA; (d) examples of diagrams that are neglected. Dashed lines: electron-impurity interaction. Thin full oriented lines: bare propagators. Thick full oriented lines: dressed propagator. Crosses represent impurities. }
\label{fig:figure3}
\end{center}
\end{figure}

The single particle Green's function of the averaged Hamiltonian is given by the Dyson equation~\cite{fetter}:
\begin{equation}
\lav\hat{G}\rav = \frac{1}{E-\hat{H}_\bot -\hat{\Sigma}}
\label{dyson}
\end{equation}
where $\hat{\Sigma}$ is the proper self-energy part that we calculate by means of the KVA and $\lav\ldots\rav$ denotes the spatial average. This approximation has been previously widely used~\cite{ando1, serre} although it is known to overestimate the width of the impurity bands~\cite{elliot}. This approximation is performed into two steps. First one isolates scattering processes $\hat{t}_i$ occurring on a single impurity:
\begin{equation}
\hat{t}_i= \hat{v}_i + \hat{v}_i  \hat{P} \hat{v}_i  + \ldots
\end{equation}
where $\hat{P} = 1/(E-\hat{H}_\bot) $ is the Green's function of the free electron. Then, one sums over all the impurities: 
\begin{equation}
\hat{t}= \sum_i \hat{t}_i= \sum_i \hat{v}_i ( \hat{1}- \hat{P}\hat{v}_i)^{-1}~,  
\label{ata}
\end{equation}
Afterwards, the bare propagator $\hat{P}$ is replaced by the dressed propagator $\hat{G}$ in the precedent equation:
\begin{equation}
\hat{t}= \sum_i \hat{v}_i ( \hat{1}- \hat{G}\hat{v}_i)^{-1} 
\label{klauder}
\end{equation}
and the average value of Eq.~\ref{dyson} $\lav \ldots \rav $ is taken over all possible positions of impurities:
\begin{equation}
\hat{\Sigma}= \lav \hat{t} \rav = \lav \sum_i   \hat{v}_i ( \hat{1}- \hat{G}\hat{v}_i)^{-1} \rav.
\label{sigma}
\end{equation}
Figure~\ref{fig:figure3}a represents Eq.~\ref{sigma} diagrammatically. Figure~\ref{fig:figure3}b represents Eq.~\ref{dyson}. Other diagrams are represented in Fig.~\ref{fig:figure3}c and Fig.~\ref{fig:figure3}d.

Although the problem has already been greatly simplified, the average is not known. To calculate the average, we follow Ando~\cite{ando1} who used two different basis. The first basis $|N,X\ket$ is used for averaging over the impurities positions. The vectors of the basis are given by
\begin{equation}
\bra\mathbf{r}|N,X\ket= 
\frac{1}{\sqrt{L}} \exp \left( i \frac{xy}{2l^2} - i \frac{Xy}{l^2} 
\right)
\xi_i(x-X) ,
\end{equation}
where $L$ is the length of the system, and $l= \sqrt{\hbar/eH}$ the radius of the cyclotron orbit. The functions  $\xi_N(x)$ are written:
\begin{equation}
\xi_N(x)= \sqrt{\frac{1}{2^NN!\sqrt{\pi} l}} \exp \left( -\frac{x^2}{2l^2} \right) H_n(x/l)
\end{equation}
and $H_n(x)$ is  the n$^{th}$ Hermite polynomial.

The second basis $|N,m\ket_{\mathbf{R}_i}$ is centered on the i$^{th}$ impurity. This is the most convenient basis for summing the different scattering processes over a given impurity because of the conservation of the cylindrical symmetry of the impurity potential. The vectors of the basis are defined by
\begin{equation}
\bra\mathbf{r}|N,m\ket_{\mathbf{R}_i} =
\bra\mathbf{R_i-r} | N,m\ket  \exp \left( i \frac{| \mathbf{r \wedge \mathbf{R}_i}| }{2l^2}  \right),
\label{phinmr}
\end{equation}
where the kets $|N,m\ket$ are defined by Eq.~\ref{phinm}. The wavefunctions of the different bases are linked to each other by the relation
\begin{equation}
|N,X\ket=
\sum_{m=-N}^{\infty} |N,m\ket_{\mathbf{R}}
\sqrt{2\pi l^2} (-1)^m  \bra\mathbf{R} | N+m, X\ket .
\label{phichange}
\end{equation}
Finally we assume that:
\begin{equation}
\lav \sum_i |\mathbf{R}_i\ket \bra\mathbf{R}_i |  \rav \simeq N_i .
\label{average}
\end{equation} 
Then, using Eq.~\ref{sigma}, \ref{phichange} and \ref{average} the self-energy part can be rewritten as:
\begin{eqnarray}
\bra N,X| \hat{\Sigma} |N',X'\ket = 2\pi l^2 N_i \sum_m \bra N,m| \hat{v} (\hat{1}- \hat{G}\hat{v})^{-1} |N,m\ket
~~\delta_{NN'} \delta_{XX'}.
\label{selfself}
\end{eqnarray}
The two main physical parameters appear in Eq.~\ref{selfself}. The first parameter is the adimensional concentration of impurities, $c= 2\pi l^2 N_i$ which represents the number of impurities seen by one electron. The second parameter is the potential $\hat{v}$. The off-diagonal elements of $\hat{v}$ are responsible for the Inter-Landau-Levels Mixing (ILLM). We have already mentioned that the potential $\hat{v}$ can be smooth or sharp depending on the location of the doping layer. In the above summation, either the potential is sharp and few values of $m$ are necessary to obtain a good convergence, or the potential is smooth and many values of $m$ are needed. This reflects the spatial extent of the eigenfunctions which need less or more basis elements for their construction.

\subsection{The density of states}
The DOS is directly related to the averaged Green's function by the formula:
\begin{equation}
\rho(E)= -\frac{1}{\pi} \Im \left( \mathrm{tr} \lav \hat{G} \rav \right),
\end{equation}
where $\lav\hat{G}\rav$ is calculated by the resolution of the self-consistent set of Eq.~\ref{dyson} and Eq.~\ref{selfself}. The calculation of the DOS consists of summing matrix (whose rank is the number of LL) indexed by the angular momentum $m$ over all values of the angular momentum $m$, in a self consistent procedure. In the previous section, the calculation of the binding energies was made by summing on the LL, for each particular value of the angular momentum $m$.\\
The DOS is represented throughout the paper by its normalized value (NDOS) defined by the relation $NDOS(E)=\rho(E)/(eB/h)$. $NDOS(E)$ is such a way that:
\begin{equation}
\int {NDOS(E)dE}=1
\end{equation}
It is represented as a function of the energy (E in $mev$ and NDOS in $mev^{-1}$) or as a function of the adimensional energy $\frac{E- \hbar  \omega_c}{\hbar \omega_c}$ (the adimensional NDOS is $ANDOS = NDOS \times \hbar \omega_c$).

\section{Density of states in presence of a smooth potential: Reference sample}
We consider first the DOS of a sample having only one $\delta$-doped layer of Si donors in the GaAlAs barrier, at a distance $z_0=-500$ $\AA$ from the interface. This is our reference sample: the disorder potential is smooth. The modification of this DOS by a strong interaction with a second layer of impurities (acceptors Be or donors Si) is discussed in the next section.\\

In this part, we have restricted our attention to the low coupling regime between LLs, when the cyclotron energy $\hbar \omega_c$ is larger than the Landau level broadening $\Gamma$. In this case, each Landau level can be treated separately. Therefore, the matrix reduces to a single real number, for each value of the angular momentum $m$. Nevertheless, a large number of m values is necessary to obtain the convergence of the self consistent calculation. The total number of $m$ necessary to reach the convergence criterium is known, when, adding a new term does not change neither the shape nor the shift of the DOS. Physically, this signifies that the smooth potential needs a large number of states $ |N,m\ket$ for its correct representation. 
\begin{figure}[htpb]
\begin{center}
\includegraphics[width =8cm]{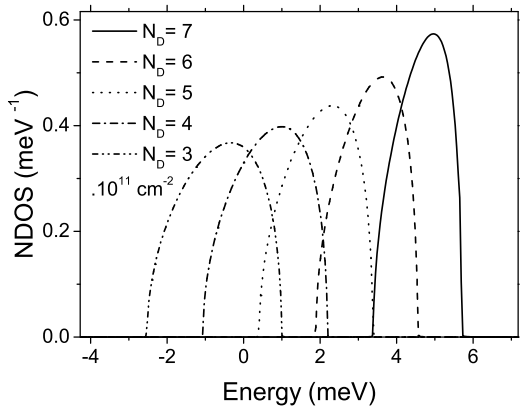}
\end{center}
\caption{Density of states at B = 10 T (ground state) for a GaAlAs/GaAs heterojunction with a $Si-\delta$-doped layer located at 500 $\AA$ from the interface. Different curves correspond to different impurity concentration $N_D = (3, 4, 5, 6).10^{11} $cm$^{-2}$. The surface density is $N_S = 3.10^{11} $cm$^{-2}$. More than hundred ``m'' have been taken into account.}
\label{fig:figure4}
\end{figure}

\begin{figure}[!h]
\begin{center}
\includegraphics[width= 8cm]{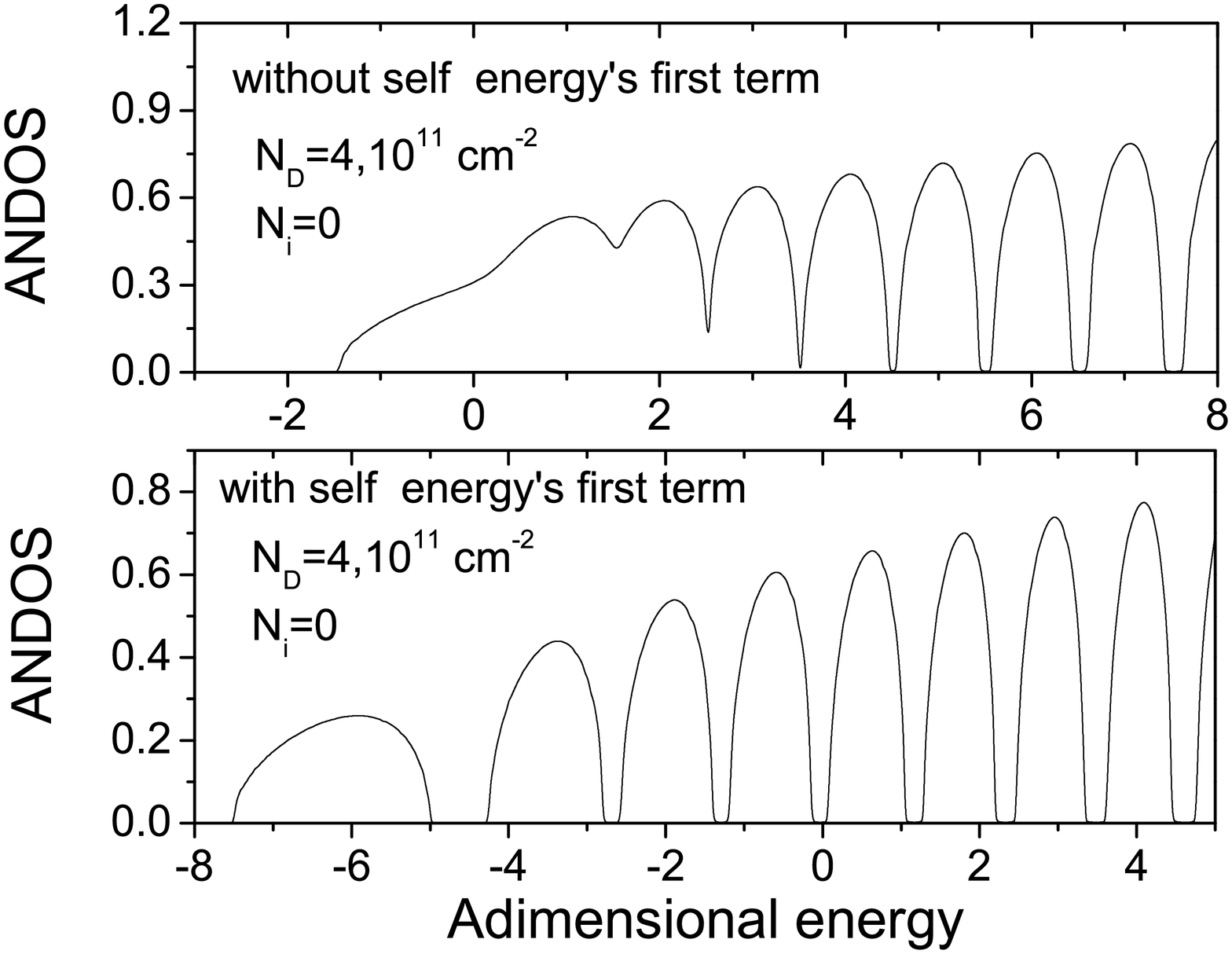}
\caption{When the first term of the self energy is taken off the equations, the Landau ladder is harmonic. When this term is kept, the Landau ladder is anharmonic.} 
\label{fig:figure5}
\end{center}
\end{figure}

Fig.4. represents the general shape of the ground Landau level whose width (overestimated by our method) is increased by the presence of impurities. There are two main differences between this shape and the elliptic shape obtained by Ando~\cite{ando1, afs, ando2} or to the gaussian shape obtained by E. Brezin et al.~\cite{brezin}. First is the asymetry of the DOS: it is in fact enlarged on its low energy side, because of the donor caracter of impurities. Second, is the global shift of the DOS towards low energy. This rigid shift is due to the first term in the calculation of the self energy, which is generally not taken into account. The shift is a decreasing function of the LL index, as illustrated in Fig.~\ref{fig:figure5}. This results in a anharmonicity of the Landau ladder. We stress that this effect should be experimentally confirmed~\cite{commentary}.  

\begin{figure}[!h]
\begin{center}
\includegraphics[width = 8cm]{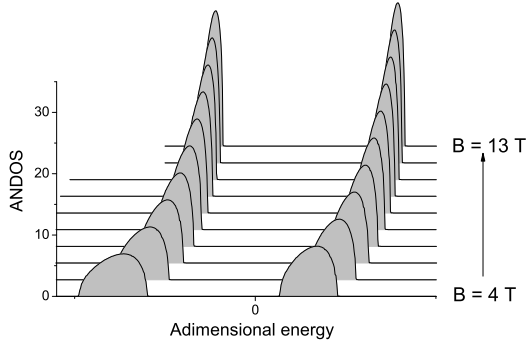}
\end{center}
\caption{Density of states for a GaAlAs/GaAs heterojunctions with only one $Si$-doped delta layer located at 500 \AA\, from the interface, for different magnetic fields. The abscissa axis is shifted by $\frac{\hbar \omega_c}{2}$ and normalized by $\hbar \omega_c$. The vertical axis is normalized by $\frac{eB}{h}/ \hbar \omega_c$.}
\label{fig:figure6}
\end{figure}

Finally, we present on Fig.~\ref{fig:figure6} the DOS of the first two LL for different magnetic field values on an adimensional scale. We observe the disparition of the shift when the magnetic field is increased. The energies of the perturbed LLs are clearly approaching the unperturbed energies at high magnetic field: the ladder is now harmonic. 
 
\section{Density of states in presence of both smooth and sharp potentials}
\subsubsection{High magnetic field}
In presence of high magnetic field, ILLM can be neglected because the adimensional impurity concentration is small ($c= 4\times 10^{-2}$ for $N_i$=$1.10^{10}cm^{-2}$ at $B$=$10T$). Therefore, the overlap integral between two different impurities becomes negligible, because the spatial spreading of the localized states shrinks with increasing $B$. Each impurity can be considered as isolated from each other and the calculations shall tend to the case of one impurity treated in ref~\cite{kubisa}.\\  

\begin{figure}[htpb]
\begin{center}
a)\includegraphics[width = 7cm]{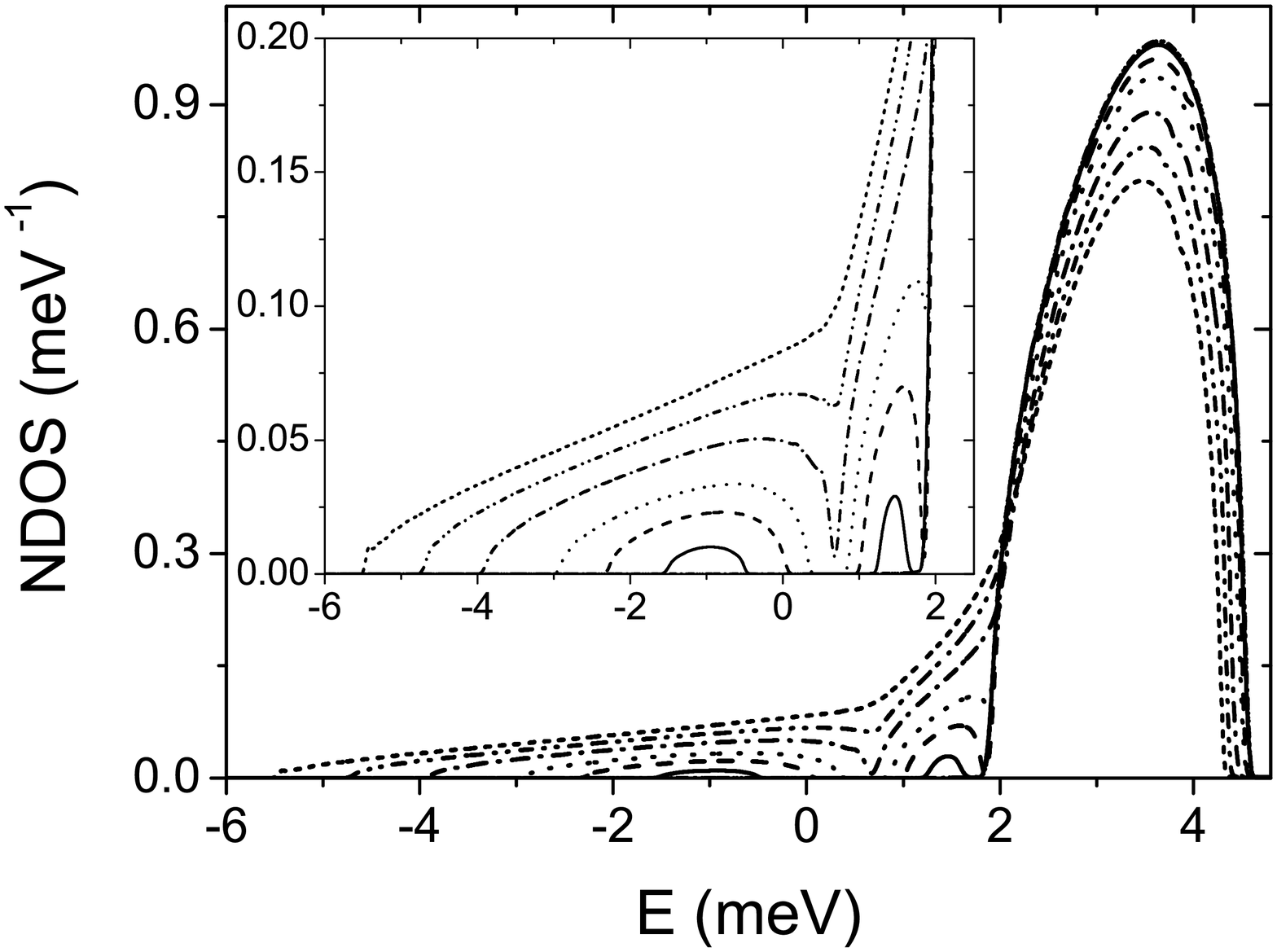}
b)\includegraphics[width = 7cm]{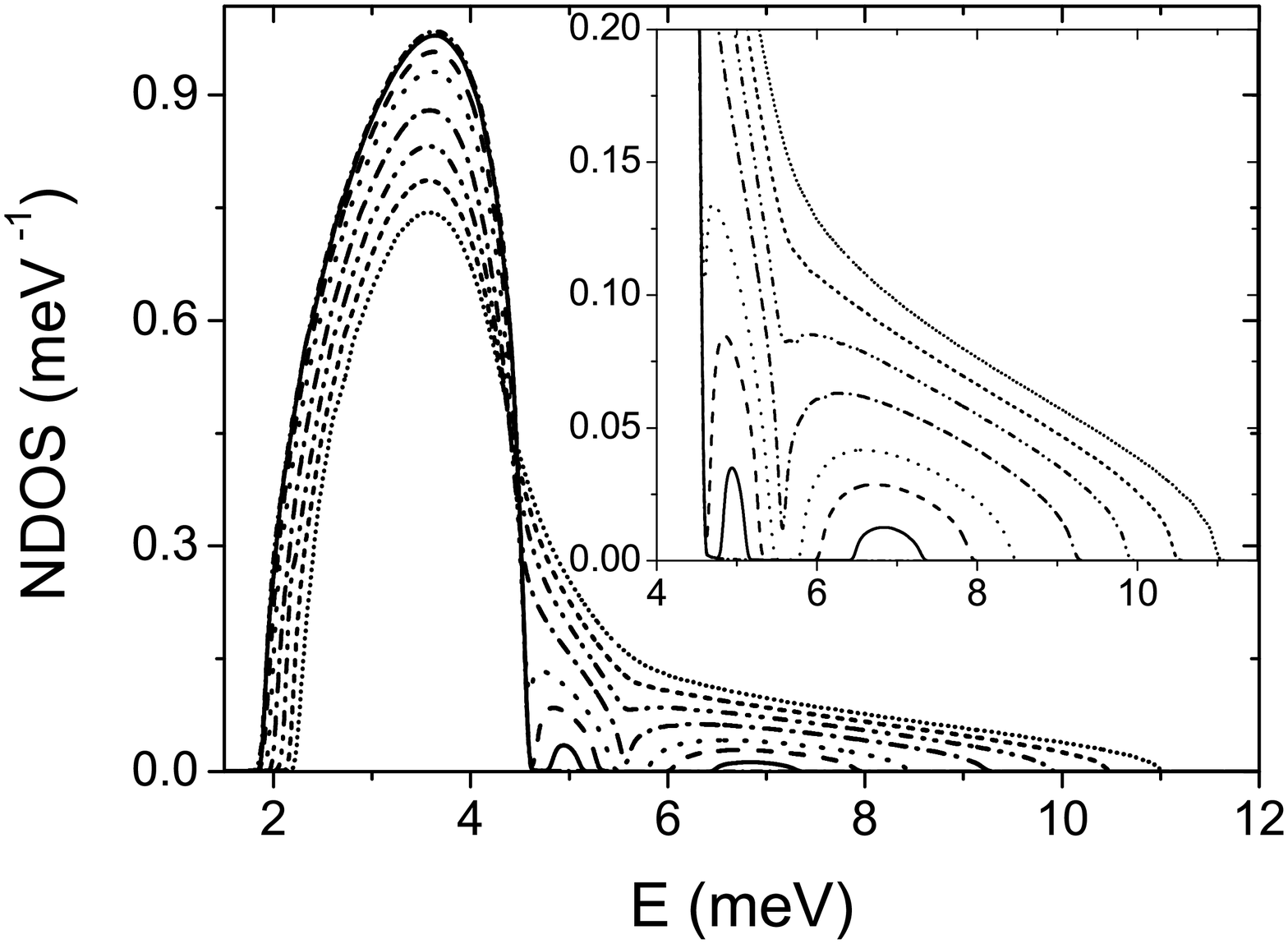}
\end{center}
\caption{Influence of the doping density on the normalized density of states at $B=10T$ for the first LL, in the case of a GaAlAs/GaAs heterojunction having two delta doped layers. First delta-doped layer: concentration $N_D = 4.10^{11} $cm$^{-2}$ of Si-atoms at $z_0=500\AA$. Second doping layer: Si atoms (curve a) or Be atoms (curve b) at $z_0=0$. Different curves correspond to different density of the second layer in the range $0.1$ $10^{10} $cm$^{-2}< N_i < 5$ $10^{10} $cm$^{-2}$. }
\label{fig:figure7}
\end{figure}

The density of states in the high field limit is represented in Fig.~\ref{fig:figure7}a and Fig.~\ref{fig:figure7}b respectively for donors and acceptors. We observe the formation of impurity bands (IB) on the low energy side in the case of donors, and on the upper energy side for acceptors. The energy ranges of IB are certainly overestimated due to the method used to treat the disorder. However, we can observe that each isolated band appearing at low density, which corresponds to levels of different angular momentum ($m=0,1,2,3,...$), is exactly centered on the binding energy calculated by Kubisa and Zawadzki~\cite{kubisa}. Counting the total number of states of an impurity band which is entirely separated from the LL, one obtains the impurity density $N_i$. If $E_{B1}$ (resp.  $E_{B2}$) represents the lower (resp. higher) band edge of the IB, then, neglecting the spin degeneracy, one obtains:\\
\[
\int\limits_{E_{B1}}^{E_{B2}} {\rho (E)dE = 2N_i}
\]
\\
We observe the formation of an impurity band for densities $N_i\leq 2.10^{10}$cm$^{-2}$. When the impurity concentration increases, the band width increases as well untill overlapping for a critical density which equals $N_{ic} = 2.10^{10}$ cm$^{-2}$ at $B=10 $ T in our case. More precisely, this critical density depends on several parameters: the position of the doping layer, the 2D electron density $N_S$, and the magnetic field B.\\  

\begin{figure}[htpb]
\begin{center}
a)\includegraphics[width = 7cm]{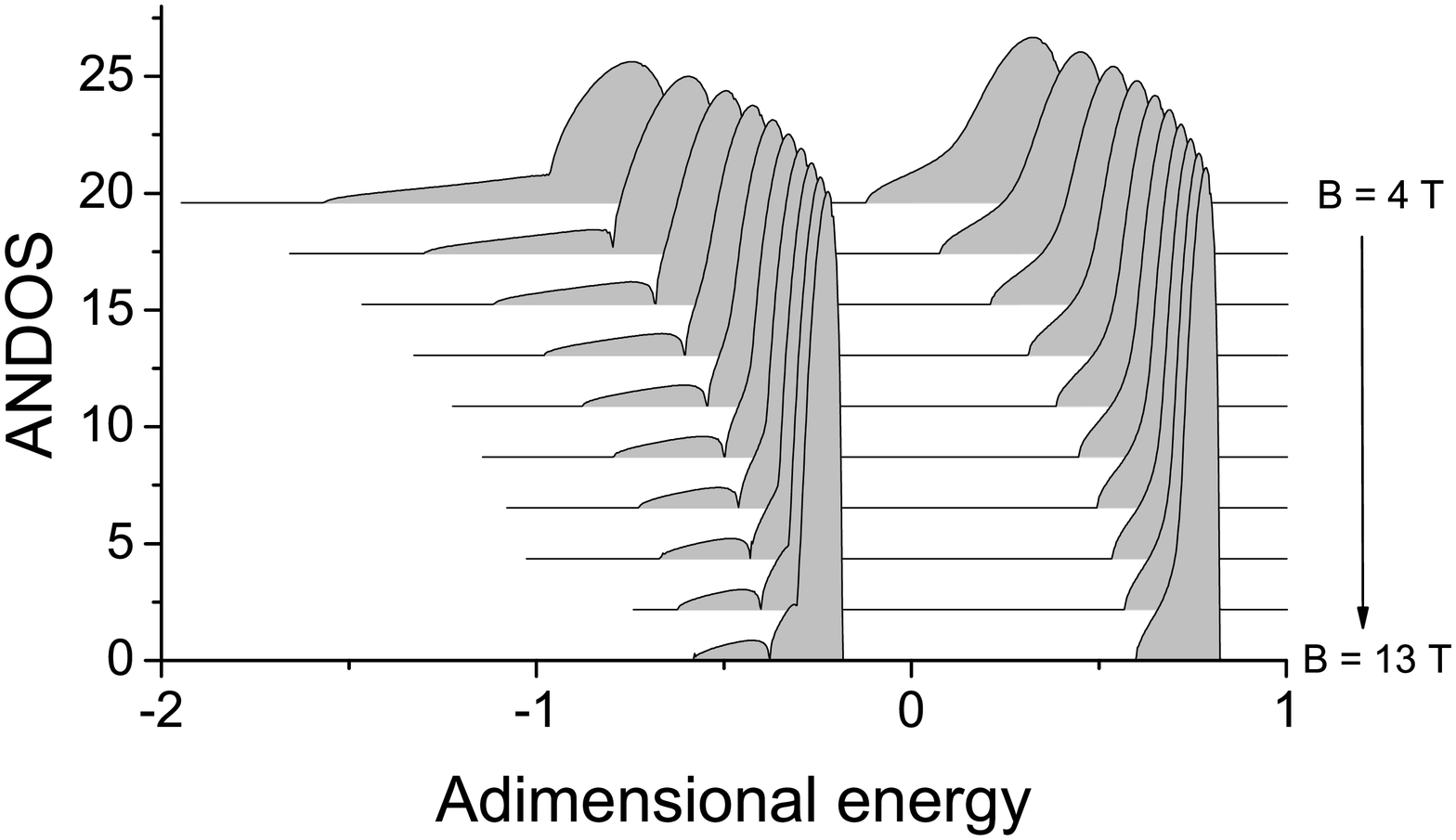}
b)\includegraphics[width = 7cm]{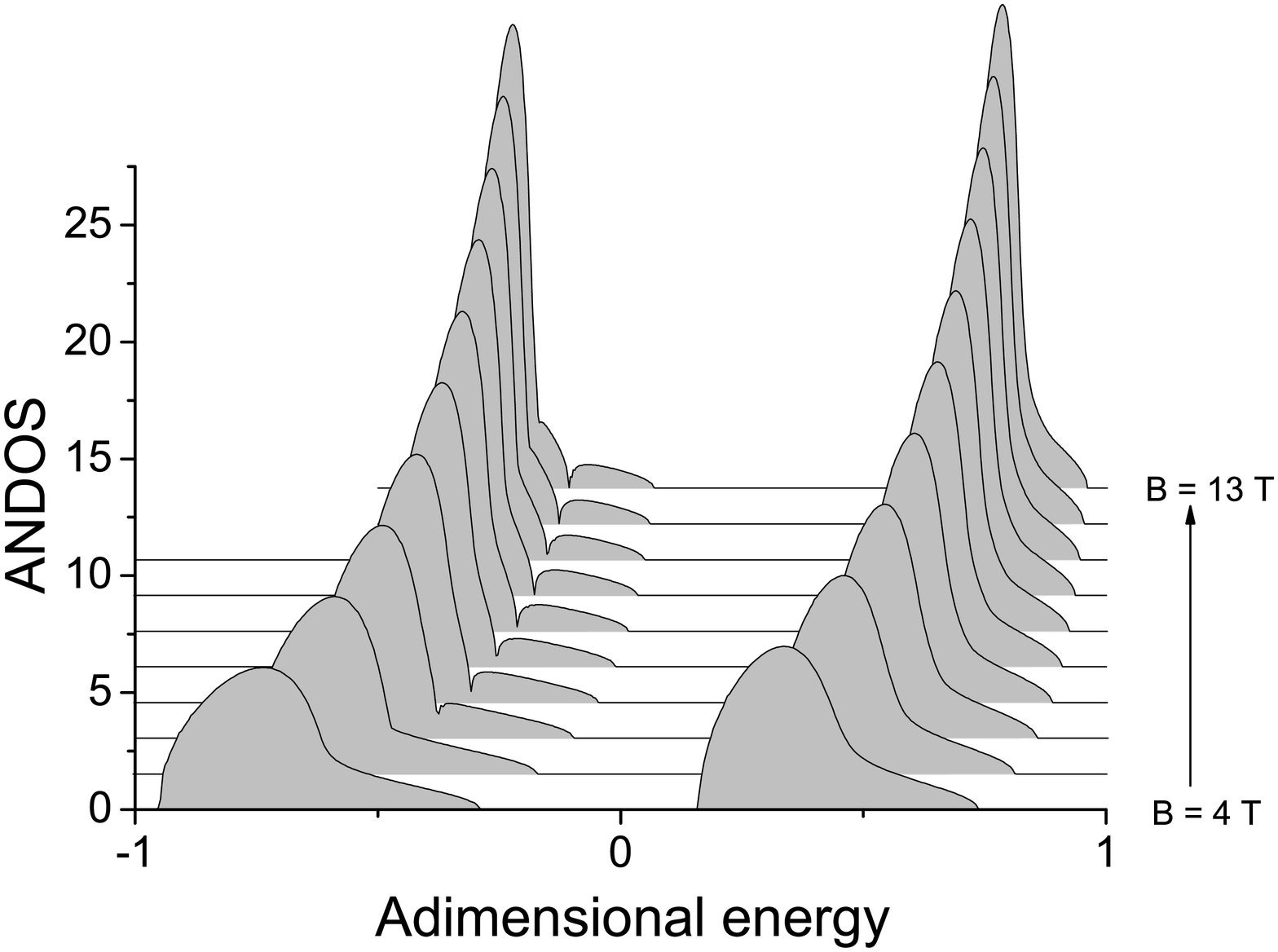}
\end{center}
\caption{Density of states for different magnetic field values, in the case of a GaAlAs/GaAs heterojunction having two delta doped layers. First delta layer: Si atoms at $z_0=500\AA$. Second doping layer: a) Si atoms at $z_0=0$ ; b) $Be$ atoms at $z_0=0$.} 
\label{fig:figure8}
\end{figure}

In Fig.~\ref{fig:figure8}, we have reported the normalized density of states for several values of the magnetic field. The two sets of curves correspond to the acceptor and donor case. We notice that the global shift represented in Fig.~\ref{fig:figure8} is not strongly modified by the additional low-density impurity layers. Then, increasing the magnetic field strength, leads to the formation of a separated impurity band for the $N=0$ LL. At low field, this impurity band collapses to form a band tail. For the first excited LL, no separated impurity band are created, because of too small binding energies. 

\subsubsection{Intermediate magnetic field}
\begin{figure}[!htpb]
\begin{center}
\includegraphics[width = 8cm]{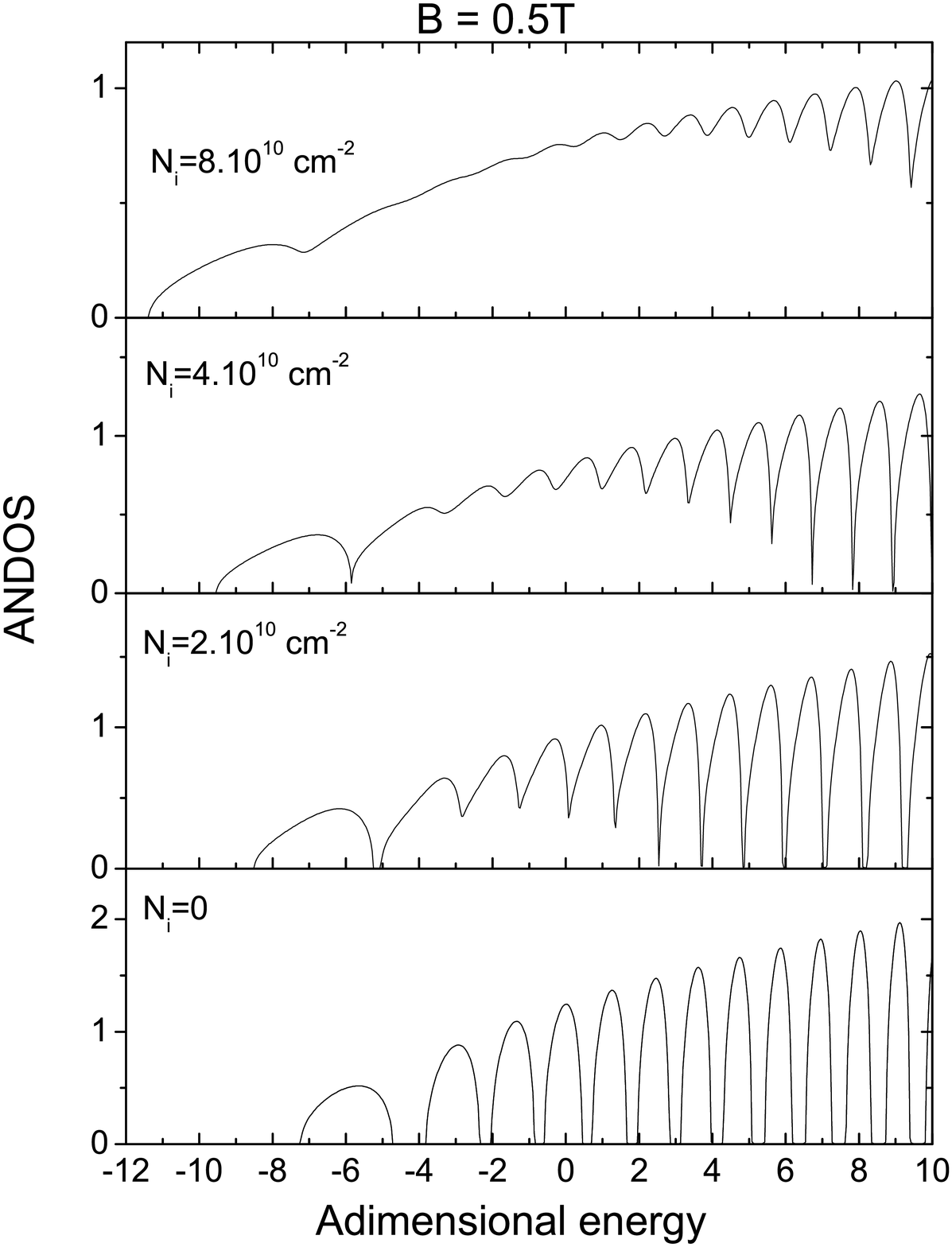}
\end{center}
\caption{Donors case: density of states for a GaAlAs/GaAs heterojunctions having a delta layer of $Si$ at 500 \AA\ from the interface, and another delta layer (density $N_i$) of $Si$ located at the interface. The magnetic field is B = 0.5 T.}
\label{fig:figure9}
\end{figure}

\begin{figure}[!htpb]
\begin{center}
\includegraphics[width = 8cm]{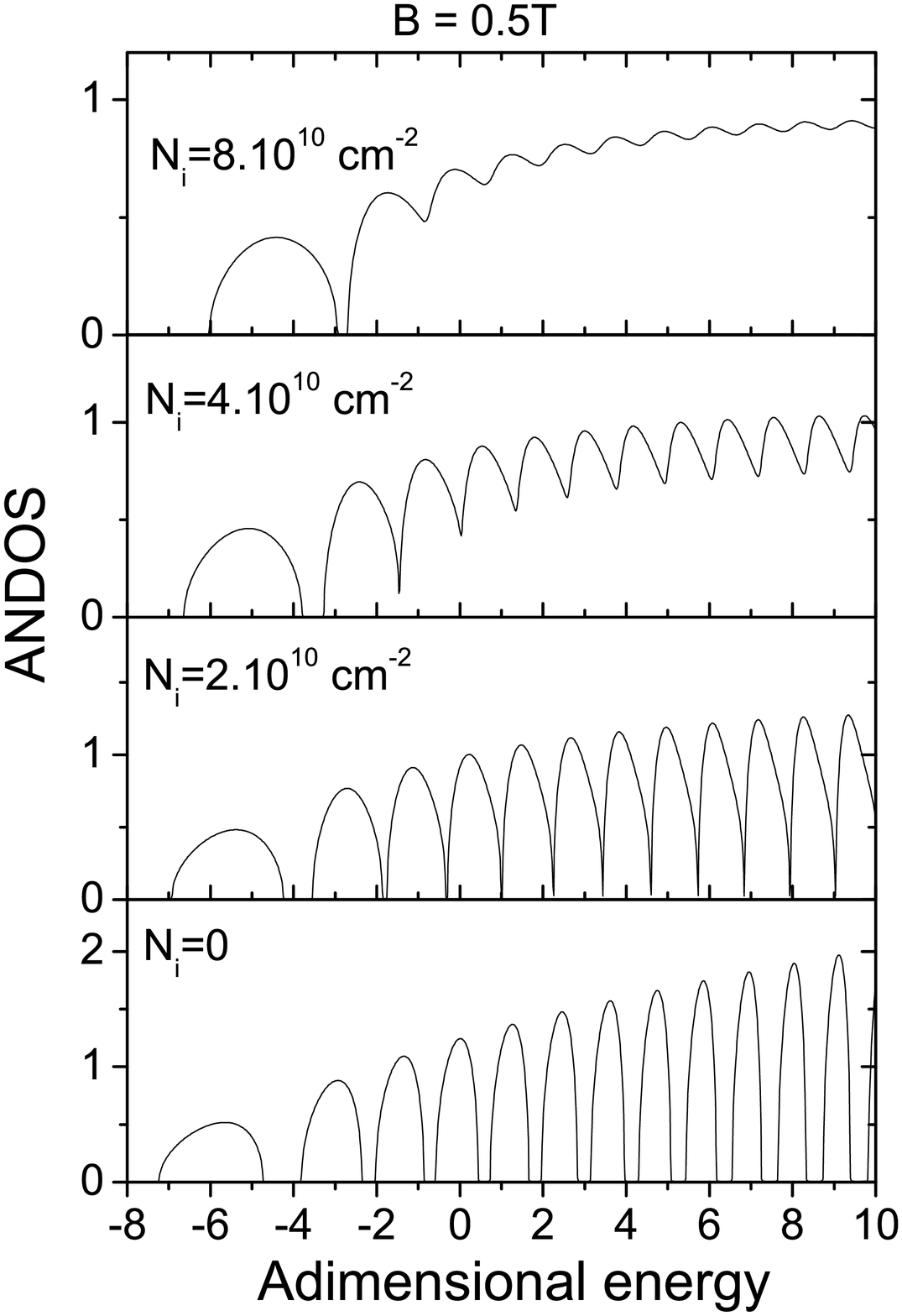}
\end{center}
\caption{Acceptors case: density of states for a GaAlAs/GaAs heterojunctions having a delta layer of $Si$ at 500 \AA\ from the interface, and another delta layer (density $N_i$) of $Be$ located at the interface. The magnetic field is B = 0.5 T.}
\label{fig:figure10} 
\end{figure}

We comment in this paragraph, the overlap of LLs as a consequence of their coupling when the parameter $c$ overpasses $1$ (by increasing $N_i$). The DOS calculated for $B=0.5$ T, is represented in the case of donors in Fig.~\ref{fig:figure9} and in the case of acceptors in Fig.~\ref{fig:figure10}. \\

In both cases the DOS is enlarged in the low energy side by the long range potential fluctuations of attractive Si parents donors at 500 \AA\ from the interface. The addition of short range potential fluctuations at the interface enlarges the DOS differently in case of acceptors or donors. Attrative Si impurities enlarge the DOS on the same side than the parent donors, whereas Be impurities enlarge the DOS on the other side.

Up to the critical value $c=1$, the degeneracy of the first LL is large enough to contain all impurities states. When the density of the impurity layer increases ( $c \geq  1$), the LLs overlap as shown in Fig.~\ref{fig:figure9}. The first LL remains separated from the higher LLs for a wider range of concentration values, because the shift is more important for low LL index $N$, as previously remarked.\\

Fig.~(\ref{fig:figure10}) is dedicated to the acceptor case. Now, the first LL remains separated from the Landau ladder even at the higher density of $N_i=8.10^{-10}cm^{-2}$ and the merging of LLs begins by the high values of $N$. This difference between acceptors and donors is related to the difference which is observed at $B=0T$ : there only exists donors states at very low B. Moreover, in the case of attractive potentials, the coupling between LLs increases the binding energies of the lowest LLs and decreases the binding energy of the highest LLs, resulting in a donor state of finite binding energy at $B=0T$ (see Fig.2).  

\subsubsection{Low magnetic field}
\begin{figure}
\begin{center}
\includegraphics[width = 8cm]{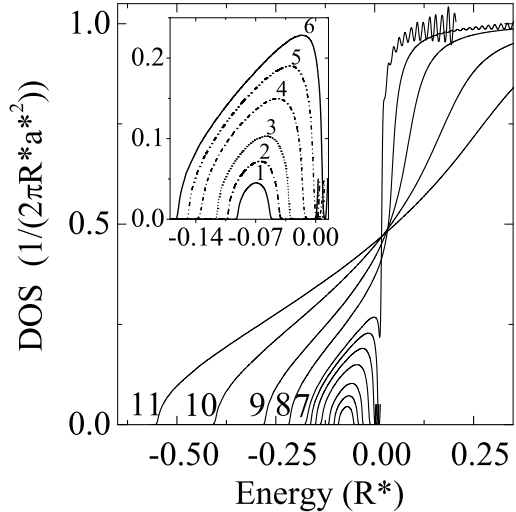}
\end{center}
\caption{The density of states in the case of magnetodonors, at $B=0.1$T, for different impurity concentrations. The curves labelled from 1 to 11 correspond to concentrations 0.01, 0.025, 0.05, 0.1, 0.15, 0.2, 0.25, 0.5, 1, 2.5 and 5.0 $\times$ 10$^{10}$cm$^{-2}$ respectively. Curve 7 corresponds to the concentration $N_i= 0.25$ 10$^{10}$cm$^{-2}$, for which $c=1$. Inset: enlargement of the IB at low donor concentrations. The flat conduction band degeneracy is equal to $\frac{m}{\pi \hbar^2}$.}
\label{fig:figure11}
\end{figure}

Fig.~\ref{fig:figure11} shows the DOS in the case of donors impurities at $B=0.1$T as a function of energy. Different values of the donor concentration $N_i$ have been considered, and the rigid energy shift of the different curves has been removed. The energy reference is fixed at the center of the first unperturbed Landau level. The high energy side of some curves has been cut off for clarity. In the acceptor case, which is not represented here, no IB below the conduction band appear . \\
At low field, ILLM becomes dominant: LLs overlap and tend to the flat Conduction Band (CB) of a 2DEG at $B=0$T. At such a low $B$, the degeneracy of one LL is smaller than $2 N_i$ and several LLs contribute to the formation of the IB (ILLM). Fig.~\ref{fig:figure11} represents the donor case. In agreement with Fig.2, only one IB exists, corresponding to $m=0$.  \\

The critical density ($c=1$) is represented by trace $7$. Above this value, the IB and the CB merge and form a band tail. At vanishing concentration $c$, the position of the IB is again in perfect agreement with the binding energies calculated in section~2 (see Fig.~\ref{fig:figure2}).

These results are in good agreements with the theoretical results of Gold, Serre and Ghazali~\cite{gold} who calculated the DOS of a 2DEG in the case $B=0T$ within the same Klauder's approximation. 

We do not observe the formation of a second IB that could originate from states of higher momenta. As a matter of fact, our potential does
not allow the formation of excited impurity bands at vanishing magnetic field because all the binding energies vanish at $B=0$T, except for $m=0$. The disappearance of the IB occurs at a concentration $N_i^c = 0.5$ $10^{10}$cm$^{-2}$. Gold~{\it et al.} pointed out that the enhanced Thomas-Fermi screening they used implied large impurity bandwith. However the critical metal-insulator densities we found are so small that they are more probably related to the errors induced by the Klauder V approximation itself~\cite{monecke}.\\

\section{Quantum Hall effect} 
It has been established~\cite{haug} in GaAs/GaAlAs heterostructures that 2DEG perturbed by acceptors or donor impurities exhibits shifts of the quantum Hall plateaus relative to the line for the classical Hall resistance. If acceptors are added, then the plateaus shift to lower filling factor $\nu$. At the contrary, if donors are added to the 2DEG, then the plateaus shift to higher $\nu$. These shifts could be succesfully explained by the asymmetry of the DOS in each LL.  Because our theoretical approach takes into account ILLM, the Hall resistivity $R_H$ at both high and low magnetic fields can be calculated and therefore a direct comparison between the classical Hall effect and the shifts of the plateaus is possible.

\begin{figure}
\begin{center}
\includegraphics[width = 8cm]{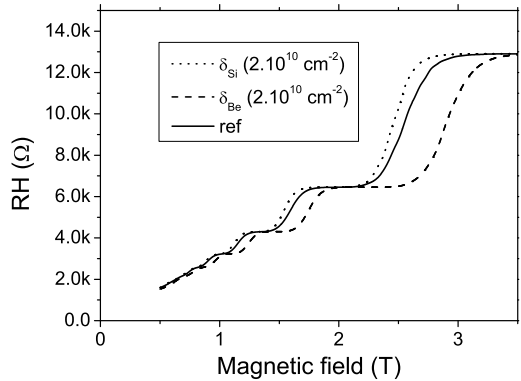}
\end{center}
\caption{Three curves corresponding to three different ``samples''. Reference is represented by a full line. The Hall resistance of the acceptor-doped sample is shifted towards low filling factors (higher B), whereas it is shifted towards high filling factors (lower B) in the case of donors.}
\label{fig:figure12}
\end{figure}

For sake of simplicity we assume that only one state per LL is delocalized whose energy $E_n$ is given by the maximum of the $n$-th LL~\cite{laughlin}. Furthermore we assume that this state exists even in the presence of strong disorder and ILLM and that it contributes to the conductivity with $e/h$. Within these approximations, the adiabatic conductivity is given by
\begin{equation}
R_H^{-1}= \frac{e^2}{h} \sum_n f (E_F-E_n)~~~,
\end{equation}
where $f$ is the Fermi distribution and $E_F$ the Fermi energy. Figure~\ref{fig:figure12} shows the calculated Hall resistance as a function of the
magnetic field $B$. For all of the three cases of Figure~\ref{fig:figure12} a $\delta$-doped layer located at $500 \AA$ from the interface, with a concentration $N_D= 4. 10^{11}$cm$^{-2}$, contains the parent donors. A second $\delta$ layer of donors (acceptors) is added at $z_0=0$, and shifts the curve towards low (high) magnetic fields, compared to the reference case. The figure reproduces well the shift of the Hall plateaus, and must be compared to Fig.~1 and Fig.~2 of Ref.~\cite{haug}. Experimentally, the shift of the plateaus is systematically much less pronounced for donors than for acceptors, and this effect is reproduced here. The reason lies in the fact that the remote layer is always doped with donors. At low field the three curves of Fig.~\ref{fig:figure12} merge into the same classical line.

\section{Conclusion}
We have used the multiple scattering approach proposed by Klauder and the averaging procedure proposed by Ando to calculate the density of states of a disordered two dimensional electron gas in the presence of a magnetic field of arbitrary strength. The impurities are randomly distributed in a layer at a given distance from the electron gas. Each impurity creates a Coulomb potential. The screening of these potentials due to the 2D electrons was taken into account. The impurities were either donors or acceptors.

At low field, in the donor case, the density of states exhibits either an impurity band or a band tail, depending on the concentration. In the acceptor case, there is neither impurity band nor band tail for low field. 

At higher field, one observes that impurity bands split from the Landau Levels. The situation between donors and acceptors becomes symmetric only at high fields, where the mixing between the Landau levels can be neglected.  
 
Our results are in qualitative agreement with  transport experiments on magnetic-field-induced metal-nonmetal transition in GaAs-GaAlAs
heterostructures~\cite{robert}, with recent magneto-photoluminescence experiments~\cite{vicente} and with transport experiments in the Quantum Hall regime~\cite{haug}. This model can be extended to calculate magneto-transport and magneto-optical properties of the 2D electron gas.


\begin{thebibliography}{99}
\bibitem{haug} R.J. Haug, R.R. Gerhardts, K. von Klitzing, K. Ploog, Phys. Rev. Lett. {\bf 59}, 1349 (1987).

\bibitem{merkt} K. Buth, M. Widmann, A. Thieme and U. Merkt, Semicond. Sci. and Technol. {\bf 18}, 434 (2003), U. Merkt, Phys. Rev. Lett. {\bf 76}, 1134 (1996), M. Widmann, U. Merkt, M. Cortes, W. Haussler, and K. Eberl, Physica B {\bf 249}, 762 (1998)

\bibitem{bonifacie} S. Bonifacie, Y. Meziani, S. Juillaguet, C. Chaubet, A. Raymond, W. Zawadzki, V. Thierry Mieg and J. Zeeman, Phys. Rev. B {\bf 68}, 165 330 (2003).

\bibitem{meulen} H.P. van der Meulen, D. Sarkar, J.M. Calleja, R. Hey, K.J. Friedland, and K. Ploog, Phys. Rev. B {\bf 70}, 155 314 (2004).
\bibitem{zhao} Q.X. Zhao, S. Wongmanerod, M. Willander, P.O. Holtz, S.M. Wang, and M. Sadeghi, Phys. Rev. B {\bf 63}, 195 317 (2001).

\bibitem{halsall} M.P. Halsall, P. Harrison, J.-P.R. Wells, I.V. Bradley, and H. Pellemans, Phys. Rev. B {\bf 63}, 155 314 (2001).

\bibitem{Wongmanerod} S. Wongmanerod, B.E. Sernelius, P.O. Holtz, B. Monemar, O. Mauritz, K. Reginski, and M. Bugajski, Phys. Rev. B {\bf 61}, 2794 (2000).

\bibitem{holtz} P.O. Holtz, A.C. Ferreira, B.E. Sernelius, A. Buyanov, B. Monemar, O. Mauritz, U. Ekenberg, M. Sundaram, K. Campman, J.L. Merz, and A.C. Gossard, Phys. Rev. B {\bf 58}, 4624 (1998).

\bibitem{hayne} M. Hayne, A. Usher, A.S. Plaut, and K. Ploog, Phys. Rev. B {\bf 50}, 17208 (1994).

\bibitem{kukushkin} I.V. Kukushkin, R.J. Haug, K. v. Klitzing, K. Eberl, and K. Totemeyer, Phys. Rev. B {\bf 50}, 11259 (1994).

\bibitem{prange} The quantum Hall effect, edited by R.~E.~Prange and M.~Girvin, Springer-Verlag, New-York (1987).


\bibitem{kubisa} M.~Kubisa and W.~Zawadzki, Semicond. Sci. Technol. {\bf 11}, 1263-1267 (1996).

\bibitem{klauder} J.~R.~Klauder, Ann. Phys. (N.Y.) {\bf 14}, 43 (1961). 
(1996). 

\bibitem{ando1} T.~Ando and Y.~Uemura, J. Phys. Soc. Jpn. {\bf 36}, 959
  (1974); T.~Ando, J. Phys. Soc. Jpn. {\bf 36}, 1521 (1974).


\bibitem{serre} J.~Serre and A.~Ghazali, Phys. Rev. B {\bf 28}, 4704
  (1983). 
          
\bibitem{gold} A.~Gold, J.~Serre and A.~Ghazali, Phys. Rev. B {\bf 37}, 4589
  (1988). 

\bibitem{monecke} J.~Monecke, J. Kortus and W.~Cordts, Phys. Rev. B {\bf 47},
  9377 (1993).

\bibitem{afs} T.~Ando, A.~B.~Fowler and F.~Stern, Rev. Mod. Physics {\bf 54}, 437 (1982).

\bibitem{joynt} R.~Joynt and R.~E.~Prange, Phys. Rev. B {\bf 29}, 3303 (1984);
  G.~F.~Giuliani, J.~J.~Quinn and S.~C.~Ying, Phys. Rev B {\bf 28}, 2969
  (1983); B.~I.~Halperin, Phys. Rev. B {\bf 25}, 2185 (1982);  D.~J.~Thouless,
  J. Phys. C {\bf 14}, 3475 (1981).

\bibitem{brezin} E.~Brezin, D.~J.~Gross and C.~Itzykson, Nuclear Physics B
{\bf 235}, 24 (1984).

\bibitem{wegner} F.~Wegner, Z. Phyzik B {\bf 51}, 279 (1983). 

\bibitem{affleck} I.~Affleck, J. Phys. C {\bf 16}, 5839 (1983).

\bibitem{ando2} T.~Ando, J. Phys. Soc. Jpn. {\bf 52}, 1740 (1983);
 T.~Ando, J. Phys. Soc. Jpn. {\bf 53}, 3101 (1984); 
T.~Ando, J. Phys. Soc. Jpn. {\bf 53}, 3126 (1984).

\bibitem{brum} J.~A.~Bastard and L.~Guillemot, Phys. Rev. B {\bf 30}, 905
  (1984). 

\bibitem{bastard} G.~Bastard, {\it Wave Mechanics Applied to Semiconductor
    Heterostructures} (Editions de Physique, Les Ulis, 1988). 

\bibitem {price} J.~Price , J. Vac. Sci. Technol. {\bf 19}, 599 (1981).

\bibitem{stern} F.~Stern, Phys. Rev. B {\bf 18}, 546 (1967). 

\bibitem{fetter} A.~L.~Fetter and J.~D.~Walecka, {\it Quantum theory of many-particle
    systems} (MacGraw-Hill Publishing Compagny, New-York, 1971).

\bibitem{elliot} R.~J.~Elliot, J.~A.~Krumhansl and P.~L.~Leath,
  Rev. Mod. Phys. {\bf 46}, 465 (1974).

\bibitem{commentary} The magneto-photoluminescence experiments are usually performed at high magnetic field, whereas the anharmonicity is enhanced at low values of the magnetic field. This is why this effect has not been experimentally evidenced. However, recent magneto-photoluminescence experiments not yet published, confirm our calculations (private comunication).

\bibitem{laughlin} R.B. Laughlin, Phys. Rev. B {\bf 23}, 5632 (1981).

\bibitem{robert} J.~L.~Robert, A.~Raymond, L.~Konczewicz, C.~Bousquet,
  W.~Zawadzki, F.~Alexandre, I.~M.~Masson, J.~P.~Andre and P.M.~Frijlink,
  Phys. Rev. B {\bf 33}, 5935 (1986).

\bibitem{vicente} P.~Vicente, A.~Raymond, M.~Kamal Saadi, R.~Couzinet,
  M.~Kubisa, W.~Wawadzki and B.~Etienne, Solid State Commun. {\bf 96}, 90
  (1995). 


\end{thebibliography}
\end{document}